\documentclass[aip, amsmath,amssymb]{revtex4-1}

\usepackage{graphicx}% Include figure files
\usepackage{dcolumn}% Align table columns on decimal point
\usepackage{bm}% bold math
\usepackage{color}% bold math

\begin{document}

\title{Electron spin resonance spectroscopy of small ensemble paramagnetic spins using a single nitrogen-vacancy center in diamond}

\author{Chathuranga Abeywardana}
\affiliation{Department of Chemistry, University of Southern California, Los Angeles CA 90089, USA}

\author{Viktor Stepanov}
\affiliation{Department of Chemistry, University of Southern California, Los Angeles CA 90089, USA}

\author{Franklin H. Cho}
\affiliation{Department of Physics and Astronomy, University of Southern California, Los Angeles CA 90089, USA}

\author{Susumu Takahashi}
\email{susumu.takahashi@usc.edu}
\affiliation{Department of Chemistry, University of Southern California, Los Angeles CA 90089, USA}
\affiliation{Department of Physics and Astronomy, University of Southern California, Los Angeles CA 90089, USA}

\date{\today}

\begin{abstract}
A nitrogen-vacancy (NV) center in diamond is a promising sensor for nanoscale magnetic sensing.
Here we report electron spin resonance (ESR) spectroscopy using a single NV center in diamond.
First, using a 230 GHz ESR spectrometer, we performed ensemble ESR of a type-Ib sample crystal
and identified a substitutional single nitrogen impurity as a major paramagnetic center in the sample crystal.
Then, we carried out free-induction decay and spin echo measurements of the single NV center
to study static and dynamic properties of nanoscale bath spins surrounding the NV center.
We also measured ESR spectrum of the bath spins using double electron-electron resonance spectroscopy with the single NV center.
The spectrum analysis of the NV-based ESR measurement identified that the detected spins are the nitrogen impurity spins.
The experiment was also performed with several other single NV centers in the diamond sample and demonstrated that the properties of the bath spins are unique to the NV centers indicating the probe of spins in the microscopic volume using NV-based ESR.
Finally, we discussed the number of spins detected by the NV-based ESR spectroscopy.
By comparing the experimental result with simulation, we estimated the number of the detected spins to be $\leq$ 50 spins.
\end{abstract}

\maketitle

\section{introduction}
A nitrogen-vacancy (NV) center is a paramagnetic defect center in diamond.
A NV center is a great testbed to investigate quantum physics because of its unique electronic, spin, and optical properties
including its stable fluorescence (FL) signals,~\cite{Gruber97}
long decoherence time,~\cite{Kennedy03, Gaebel06, Childress06, Balasubramanian09, Takahashi08}
and capability to initialize the spin states of NV centers by applying optical excitation and to readout the states by measuring the FL intensity.~\cite{Jelezko04}
A NV center is also a promising magnetic sensor~\cite{Degen08, Balasubramanian08, Maze08, Taylor08, Steinert13, Kaufmann13, Mamin13sci, Staudacher13, Ohashi13, muller14, Maletinsky12} because its extreme sensitivity to the surrounding electron~\cite{Gaebel06, Hanson06, Hanson08} and nuclear spins.~\cite{Childress06, Balasubramanian09}
Spin sensitivity of electron spin resonance (ESR) spectroscopy is drastically improved using NV centers.
ESR spectrum of small ensemble electron spins has been measured using a double electron-electron resonance (DEER) spectroscopy with single NV centers.~\cite{deLange12, Mamin12, Laraoui12, Knowles14}
ESR detection of a single electron spin has also been demonstrated using the DEER technique.~\cite{Grinolds13, Sushkov14, Shi15}
In addition, bio-compatibility and excellent chemical/mechanical stability of diamond makes a NV center suitable for applications of nanoscale magnetic sensing and magnetic resonance spectroscopy in biological systems.~\cite{Hall00, McGuinness11, LeSage13, Shi15}

In this article, we discuss ESR spectroscopy of a small ensemble of electron spins using single NV centers in diamond.
Although the state-of-the-art of NV-based ESR spectroscopy is the detection of a single electron spin,
the small ensemble measurement is often advantageous for applications of NV-based ESR because of less sophisticated sample preparation ({\it e.g.} higher density of the target spins which increases the coupling to the NV) and more sensitive detection ({\it e.g.} more pronounced changes in the coherence ($T_2$) and population ($T_1$) decays),
and still enables to probe nanoscale local environments which may be different from the bulk properties.
On the other hand, it is challenging to estimate the detected number of spins from the small ensemble measurement.

In the investigation, our experiment is performed with a type-Ib diamond crystal at room temperature.
Using a 230 GHz ESR spectrometer, we first perform ensemble ESR of the sample crystal and find that a major paramagnetic impurity in the sample crystal is a substitutional single nitrogen center (N spin; also known as P1 center).
Next, we detect a single NV center using FL autocorrelation and optically detected magnetic resonance (ODMR) measurements.
We then carry out Rabi, free-induction decay (FID) and spin echo (SE) measurements of the single NV center to study static and dynamic properties of bath spins surrounding the NV center.
We also employ DEER spectroscopy to measure ESR signals of the bath spins using the NV center.
The detected bath spins are identified as N spins from the analysis of the observed ESR spectrum.
Based on the intensity of the observed NV-based ESR signal, the detected magnetic field by the DEER measurement is $\sim$ 6 $\mu$T (equivalent to the magnetic field caused by a single $S=1/2$ spin with the distance of $\sim$ 7 nm).
We also study several single NV centers in the same crystal and find heterogeneity in their spin properties which
indicates that our ensemble measurement still proves bath spins in the microscopic scale.
Finally, we introduce a computational simulation method for the estimate of the number of the detected spins.
By comparing the experimental result with simulation, we estimate the number of the detected spins to be $\leq$ 50 spins.

\section{results and discussion}
\begin{figure}
\includegraphics[width=100mm]{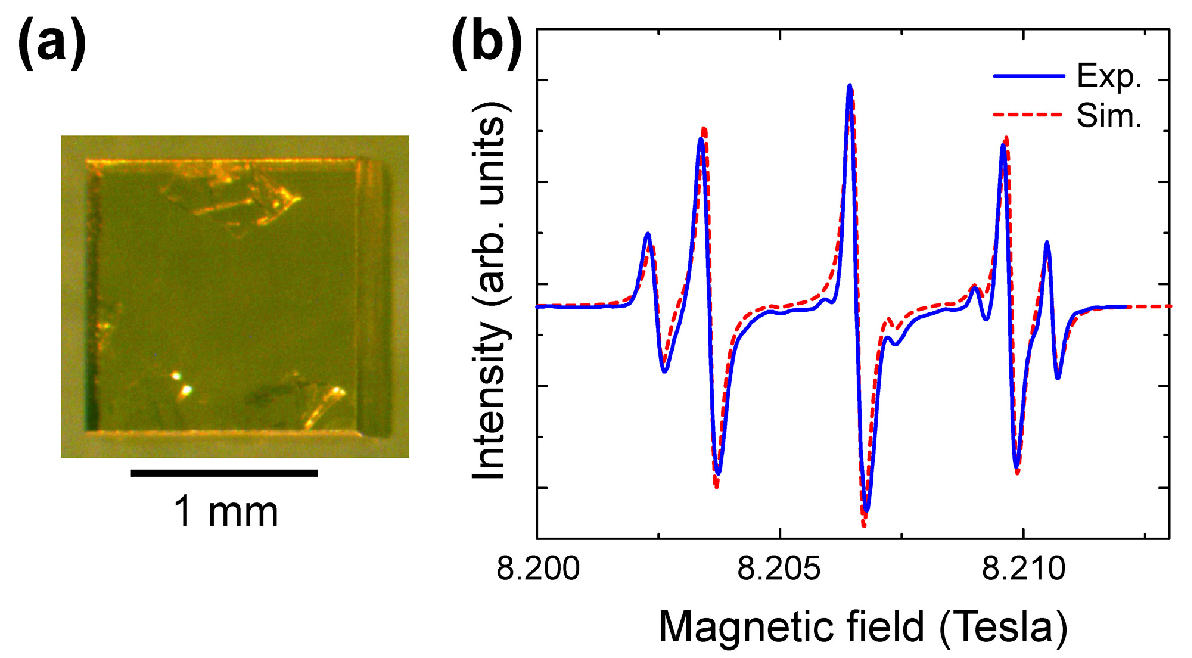}
\caption{
\label{fig:ESR}
(a) Picture of the type-Ib diamond crystal used in the investigation. (b) cw ESR spectrum of N spins measured using the 230 GHz ESR spectrometer. The spectrum was obtained by a single scan at 0.2 mT/s with field modulation of 0.03 mT at 20 kHz.
}
\end{figure}
We studied a single crystal of high-temperature high-pressure type-Ib diamond, which is commercially available from Sumitomo electric industries.
The size of the diamond crystal is $1.5\times1.5\times1$ mm$^3$ (see Fig.~\ref{fig:ESR}(a) inset).
The concentration of N spins is 10 to 100 ppm, corresponding to $4 \times 10^{15}$ to $4 \times 10^{16}$ N spins existing in diamond.
First, using a 230 GHz ESR spectrometer,~\cite{Cho14} we measured ensemble ESR of the sample diamond crystal at room temperature to characterize its bulk properties where the magnetic field was applied along the $\langle$111$\rangle$-direction of the single crystal diamond in the measurement.
As shown in Fig.~\ref{fig:ESR}(b), continuous-wave (cw) ESR spectroscopy revealed five-line ESR signals corresponding to the N spins.
The ESR intensity of the N spins was drastically stronger than the remaining signals, which indicates that the number of N spins dominates the spin population in the sample.
Moreover, no ESR signals from NV centers and other paramagnetic impurities were observed in the ESR measurement because of their low concentration in the sample crystal.
The spin Hamiltonian of N spin is given by
\begin{equation}
\label{eq:SpinH}
H_N=\mu_{B}g^{N}{\bm S^{N}}\cdot{\bm B_0}+{\bm S^{N}}\cdot\stackrel{\leftrightarrow}{A^N}\cdot{\bm I^{N}} - \mu_{n}^{N}g_{n}^{N}{\bm I^{N}}\cdot{\bm B_0} + P_z^{N} (I^N_z)^2,
\end{equation}
where $\mu_B$ is the Bohr magneton, ${\bm B_0}$ is the external magnetic field, $g^N_{x,y}$ and $g^N_z$ are the $g$-values of the N electron spin, and ${\bm S^{N}}$ and ${\bm I^{N}}$ are the electronic and nuclear spin operators, respectively.
$\stackrel{\leftrightarrow}{A^N}$ is the anisotropic hyperfine coupling tensor to $^{14}N$ nuclear spin.
The gyromagnetic ratio of $^{14}N$ nuclear spin ($\mu_{n}^{N}g_{n}^{N}/h$) is 3.077 MHz where $h$ is the Planck constant.
The last term of the Hamiltonian is the nuclear quadrupole couplings.
The previous studies measured $g^N_{x,y}=2.0024$, $g^N_z=2.0025$, $A^N_x=A^N_y=82$ MHz, $A^N_z=114$ MHz and $P_z^{N}=-5.6$ MHz.~\cite{Loubser78, Smith59, Takahashi08}
As shown in Fig.~\ref{fig:ESR}(b), the experimental data agree well with simulated ESR spectrum using Eq.~(\ref{eq:SpinH}) and the previously determined parameters.

\begin{figure}
\includegraphics[width=120mm]{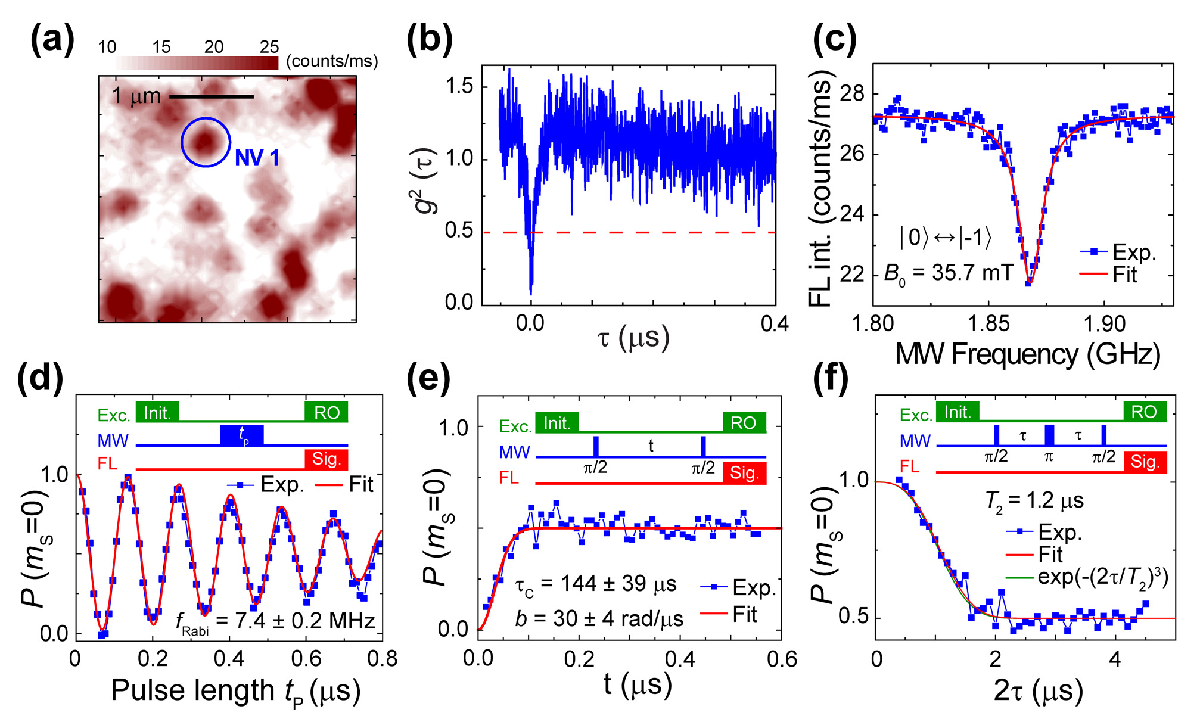}
\caption{
\label{fig:ODMR}
ODMR experiment of NV~1.
(a) FL image of the diamond crystal. The scanned area is 3$\times$3 $\mu$m$^2$.
The color scheme for FL intensity is shown in the legend.
NV~1 is circled.
(b) the autocorrelation data of NV~1.
(c) cw ODMR experiment.
The solid line indicates a fit to the Lorentzian function.
(d) Rabi oscillation experiment at $B_0=35.7$ mT and the microwave frequency of 1.868 GHz.
Square dots connected with lines indicate the measurement and solid line indicates the fit.
The oscillation frequency ($f_{Rabi}$) of 7.4 $\pm$ 0.2 MHz and the decay time ($T_d$) of 0.77 $\pm$ 0.04 $\mu$s were obtained from a fit to $1/2[\cos(2\pi f_{Rabi}t_P)\exp(-(t_P/T_d)^2)+1]$.
The inset shows the pulse sequence with excitation laser (Exc.), microwave (MW) and FL measurement (FL).
(e) FID measurement at $B_0=35.7$ mT and 1.868 GHz. The length of $\pi$/2-pulse was 34.
FL intensity decay was monitored as a function of $t$.
The inset shows the pulse sequence used in the measurement.
(f) SE experiment at $B_0=35.7$ mT and 1.868 GHz.
The lengths of $\pi$/2- and $\pi$-pulse used were 34 and 68 ns, respectively.
FL intensity decay was monitored as a function of $2\tau$.
The inset shows the pulse sequence used in the measurement.
$b$ and $\tau_C$ were extracted by fitting FID and SE data with Eqs.~(\ref{eq:FID}) and (\ref{eq:SE}).
For pulsed ODMR measurements, laser initialization pulse (Init.) of 2 $\mu$s, and laser read-out pulse (RO) and FL measurement pulse (Sig.) of 300 ns were used.
Also the FL intensity was normalized and re-scaled into the $m_S=0$ state population of NV~1.}
\end{figure}
Figure~\ref{fig:ODMR} shows ODMR measurements of a single NV center in the diamond crystal.
The ODMR experiment was performed using a home-built confocal microscope system.~\cite{Abeywardana14}
For microwave excitations, two microwave synthesizers, a power combiner, and a 10 watt amplifier were connected to a 20 $\mu$m-thin gold wire placed on a surface of the diamond sample.
First a FL image of the diamond was recorded in order to map out FL signals in the diamond crystal (see Fig.~\ref{fig:ODMR}(a)).
After choosing an isolated FL spot, we carried out the autocorrelation and cw ODMR measurements in order to identify FL signals from a single NV center.
As shown in Fig.~\ref{fig:ODMR}(b), the autocorrelation measurement of the chosen FL spot revealed the dip at zero delay which confirmed the FL signals originated from a single quantum emitter.
In addition, cw ODMR measurement of the selected single NV center was performed with application of the external magnetic field ($B_0$) of 35.7 mT along the $\langle$111$\rangle$ axis.
As shown in Fig.~\ref{fig:ODMR}(c), the reduction of the FL intensity was observed at the microwave frequency of 1.868 GHz corresponding to the ODMR signal of the $m_S=0 \leftrightarrow -1$ transition of the NV center.
Thus, the observations of the autocorrelation and cw ODMR signals confirmed the successful identification of the single NV center (denoted as NV~1).

Next, we performed pulsed ODMR measurements of NV~1.
In the pulsed measurements, a NV center was first prepared in the $m_S=0$ state by applying an initialization laser pulse,
and the microwave pulse sequence was applied for the desired manipulation of the spin state of the NV center,
then the final spin state was determined by applying a read-out laser pulse and measuring the FL intensity.
In addition, the FL signal intensity was mapped into the population of the NV's $m_S=0$ state ($P(m_S=0)$) using two references (the maximum and minimum FL intensities corresponding to the $m_S=0$ and $m_S=-1$ states, respectively).~\cite{Delange10}
Moreover, the pulse sequence was averaged on the order of $10^6$-$10^7$ times to obtain a single data point.
First, the Rabi oscillation measurement was performed at $B_0=35.7$ mT and 1.868 GHz, which corresponds to the $m_S=0 \leftrightarrow -1$ transition of NV~1.
As the microwave pulse length ($t_P$) was varied in the measurement, pronounced oscillations of $P(m_S=0)$ was observed as shown in Fig.~\ref{fig:ODMR}(c).
By fitting the observed Rabi oscillations to the sinusoidal function with the Gaussian decay envelope,~\cite{Hanson08} the lengths of $\pi/2$- and $\pi$-pulses for NV~1 were determined to be 34 and 68 ns, respectively.
Second, FID was measured using the Ramsey fringes.
The pulse sequence are shown in the inset of Fig.~\ref{fig:ODMR}(e).
As seen in Fig.~\ref{fig:ODMR}(e), the FL intensity of NV~1 was recorded as a function of free evolution time ($t$)
and FID was observed in the range of $t \leq 100$ ns.
Third, we carried out the SE measurement.
Figure~\ref{fig:ODMR}(f) shows the FL intensity as a function of free evolution time (2$\tau$).
As shown in the inset of Fig.~\ref{fig:ODMR}(f), the applied microwave pulse sequence consists of the conventional spin echo sequence, widely used in ESR spectroscopy,~\cite{Schweiger} and an additional $\pi/2$-pulse at the end of the sequence to convert the resultant coherence of a NV center into the population of the $m_S=0$ state.\cite{Jelezko04}
The FID and SE decay in electron spin baths have been successfully described by treating the bath as a classical noise field ($B_n(t)$) where $B_n(t)$ was modeled by the Ornstein-Uhlenbeck (O-U) process with the correlation function $C(t) = \langle B_n(0) B_n(t) \rangle = b^2\exp(-|t|/\tau_C)$, where the spin-bath coupling constant ($b$) and the rate of the spin flip-flop process between the bath spins ($1/\tau_C$).~\cite{Klauderanderson, Hanson08}
The FID and SE decay due to the O-U process are given by,
\begin{eqnarray}
\label{eq:FID}
FID(t) = \frac{1}{2} - \frac{1}{6} [1+2 \cos(2\pi A^{NV}_z t)]\exp[-(b \tau_C)^2(\frac{t}{\tau_C}+e^{-t/\tau_C}-1)],
\end{eqnarray}
and
\begin{eqnarray}
\label{eq:SE}
SE(t) = \frac{1}{2} + \frac{1}{2}\exp[-(b \tau_C)^2(\frac{2\tau}{\tau_C}-3-e^{-(2\tau)/\tau_C}+4e^{-(2\tau)/(2\tau_C)})],
\end{eqnarray}
where $A^{NV}_z$ = 2.3 MHz is the hyperfine coupling of NV center.~\cite{Loubser78}
In the quasi-static limit ($b\tau_C \gg 1$) indicating slow bath dynamics, $SE(t) \sim \exp[-b^2t^3/(12\tau_C)]=\exp[-(t/T_2)^3]$ where $T_2$ is the spin decoherence time.~\cite{Klauderanderson, Hanson08, Wang13}
By fitting both FID and SE data (Fig.~\ref{fig:ODMR}(e) and (f)) simultaneously with Eqs.(\ref{eq:FID}) and (\ref{eq:SE}),
we obtained $b$ and $\tau_C$ to be 30 $\pm$ 4 (rad/$\mu s$) and 144 $\pm$ 39 $\mu$s, respectively.
The result indicates that the surrounding spin bath is in the quasistatic limit ($b \tau_C = 4320 \gg 1$), therefore, $T_2$ ($=(12\tau_c/b^2)^{1/3}$) of NV~1 is 1.2 $\mu$s.
Using the previous study,~\cite{Wang13} the local concentration of the bath spins around NV~1 is estimated to be 20 ppm.

\begin{figure}
\includegraphics[width=50 mm]{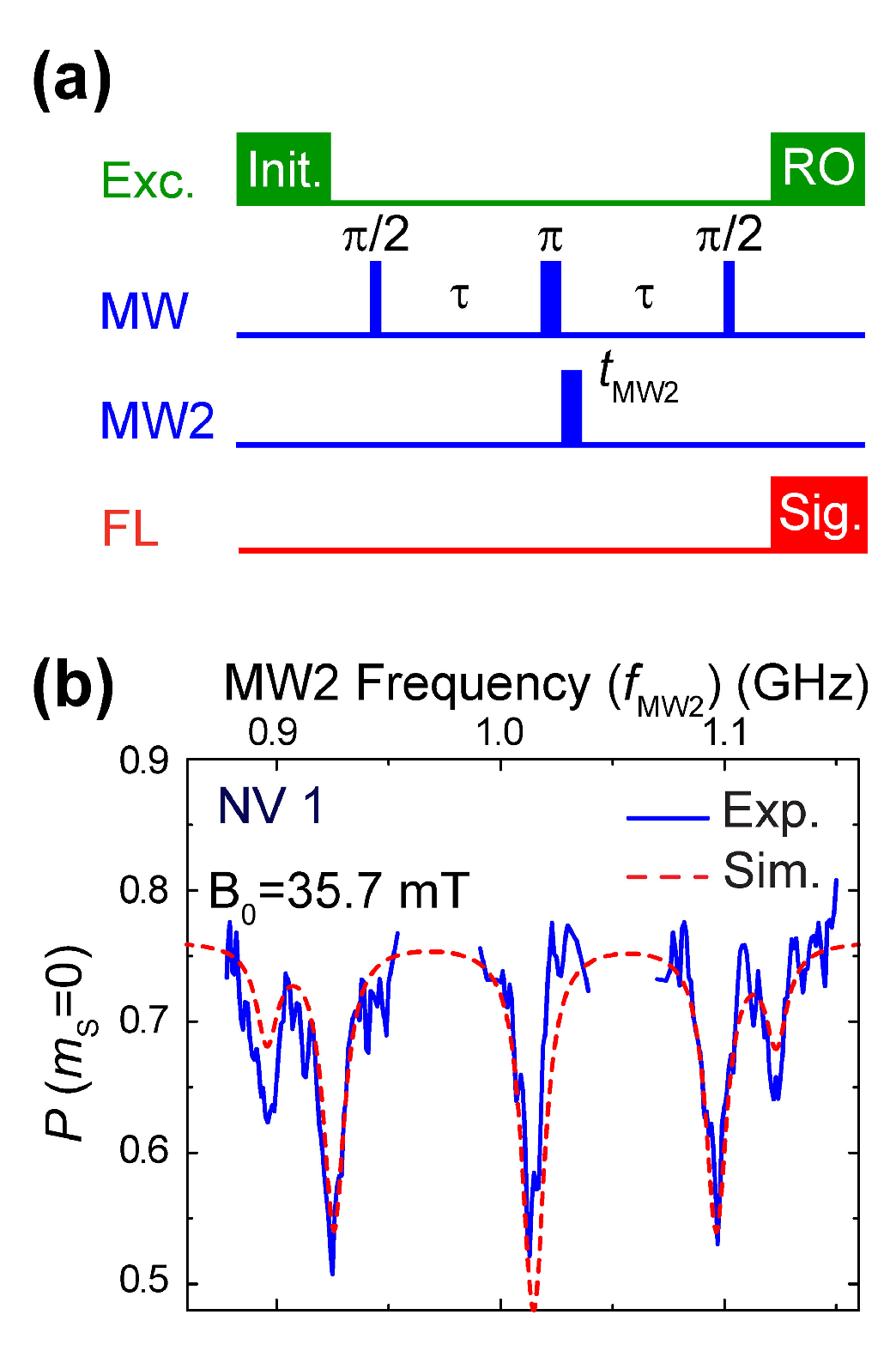}
\caption{
\label{fig:DEER}
NV-based ESR experiment using DEER spectroscopy. (a) The pulse sequence of the DEER spectroscopy.
In addition to the spin echo sequence of the NV center (denoted as MW), another microwave pulse (denoted as MW2) was applied 12 ns after the $\pi$-pulse in MW to rotate the surrounding spins.
(b) The obtained ESR spectrum.
In the measurement, $\tau =$ 500 ns and $t_{MW2}=90$ ns were used.
The simulated ESR spectrum is shown in the dotted line.
}
\end{figure}
Furthermore, we performed NV-based ESR using DEER spectroscopy at $B_0=35.7$ mT ($f_{MW1}=$1.868 GHz corresponding to the $m_S=0 \leftrightarrow -1$ transition).
Figure~\ref{fig:DEER}(a) shows the pulse sequence used in the DEER measurements,\cite{deLange12, Mamin12, Laraoui12} which was adopted from the field of ESR.~\cite{Schweiger}
The DEER sequence consists of the spin echo sequence for a NV center and an additional $\pi$-pulse at different microwave frequency (denoted as MW2 in Fig.~\ref{fig:DEER}(a)).
When the additional $\pi$-pulse is resonant with surrounding electron spins near the NV center, the magnetic moments of the surrounding spins are flipped, and this alters the magnetic dipole field experienced by the NV center from the surrounding spins during the second half of the spin echo sequence.
The alteration causes a shift in the Larmor frequency of the NV center, which leads to different phase accumulation of the Larmor precession during the second half of the sequence from the first half.
As a result, the NV center suffers phase-shift in the echo signal and reduction of FL intensity from the original spin echo signal is observed.
As shown in Fig.~\ref{fig:DEER}(b), we monitored the spin echo intensity of NV~1 at a fixed $\tau$ as a function of $f_{MW2}$, and observed clear intensity reductions at five frequencies ($f_{MW2} =$ 0.90, 0.92, 1.01, 1.10, and 1.12 GHz).
In the measurement, $\tau=500$ ns and $t_{MW2}=90$ ns were chosen to maximize the NV-based ESR signals.
As shown in Fig.~\ref{fig:DEER}(b), the resonant frequencies of the observed NV-based ESR signals were in a good agreement with the ESR of N spins calculated from Eq.~(\ref{eq:SpinH}), which confirms the observation of N spin ESR spectrum.
The DEER intensity ($I_{DEER}$) represents the change of the SE intensity given by the change in the effective dipolar field ($B_{dip,eff}$).
Assuming that $\pi$-flip of the N spins is instantaneous ({\it i.e.} the effect of a finite width of the pulse was not considered), $I_{DEER}$ = $\cos(g_{NV}\mu_B (2\tau) B_{dip,eff}/\hbar)$ where $g_{NV}=2.0028$ is the $g$-value of the NV center,~\cite{Loubser78} $\hbar$ is the reduced Planck constant, and $\tau=500$ ns in the present case.
In addition, including the effects of $T_2$ decay, the NV-based ESR signal in $P(m_S=0)$ is given by,
\begin{eqnarray}
\label{eq:NV_ESR}
I_{NV ESR} = 1/2(1+ I_{DEER} SE(2\tau)).
\end{eqnarray}
Using Eq.~(\ref{eq:NV_ESR}), we obtained that the intensity of the NV-based ESR at $f_{MW2}=0.90$ GHz ($I_{NVESR} \sim 0.625$) corresponds to $B_{dip,eff}$ $\sim$ 6 $\mu$T.
This magnetic field is equivalent to the dipole magnetic field from a $S=1/2$ single spin with the NV-spin distance ($d$) of $\sim$7 nm ($B_{dip}=\mu_0/(4 \pi)(3{\bf n}({\bf n}\cdot{\bf m})-{\bf m})/d^3)$ with ${\bf n} // {\bf m}$, $|{\bf m}| = g \mu_B m_S$, $g$ = 2 and $m_S$ = 1/2).

\begin{figure}
\includegraphics[width=120 mm]{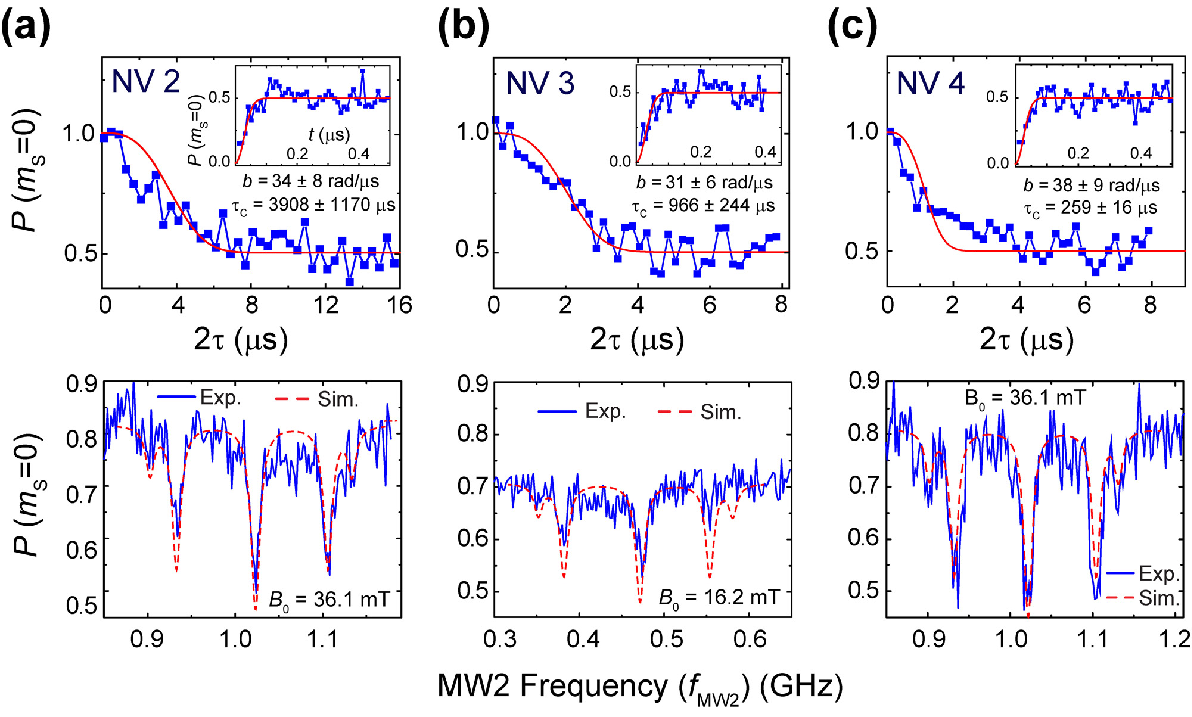}
\caption{
\label{fig:otherNVs}
SE, FID and NV-based ESR data of other three single NV centers: (a)NV 2; (b) NV 3; (c) NV 4.
}
\end{figure}
In addition to the investigation of NV~1, we have studies other NVs in the same diamond crystal.
Figure~\ref{fig:otherNVs} shows the set of FID, SE and DEER data from other three NV centers (labeled as NV~2-4).
The results from NV~1-4 were different because of their heterogeneous nanoscale local environments.
As shown in Fig.~\ref{fig:otherNVs}(a)-(c), we found that the obtained $\tau_C$ values were varied widely from 144 $\mu$s to 3908 $\mu$s.
On the other hand, the variation of $b$ = 30-38 rad/$\mu s$ is much smaller than $\tau_C$.
As the result, $T_2$ of NV~1-4 ranges from 1.2 to 3.4 $\mu$s.
We also noticed that all NV~1-4 are in the quasistatic limit ($b \tau_C \gg 1$).
As shown in Fig.~\ref{fig:DEER}(b) and Fig.~\ref{fig:otherNVs}, the NV-based ESR spectra of NV~1-4 were also quite different.
The NV-based ESR spectra from NV~1, 2 and 4 displays five-line ESR signals (Fig.~\ref{fig:DEER}(b) and Fig.~\ref{fig:otherNVs}(a)(c))
whereas NV 3 only shows three visible signals at 0.381 0.474 and 0.553 GHz (Fig.~\ref{fig:otherNVs}(b)).
Possible reasons of the difference are heterogeneity of the number and the spatial configuration of N spins around the NV centers,
and an uneven distribution of the N spin orientations due to a small number of N bath spins.

\begin{figure}
\includegraphics[width=100 mm]{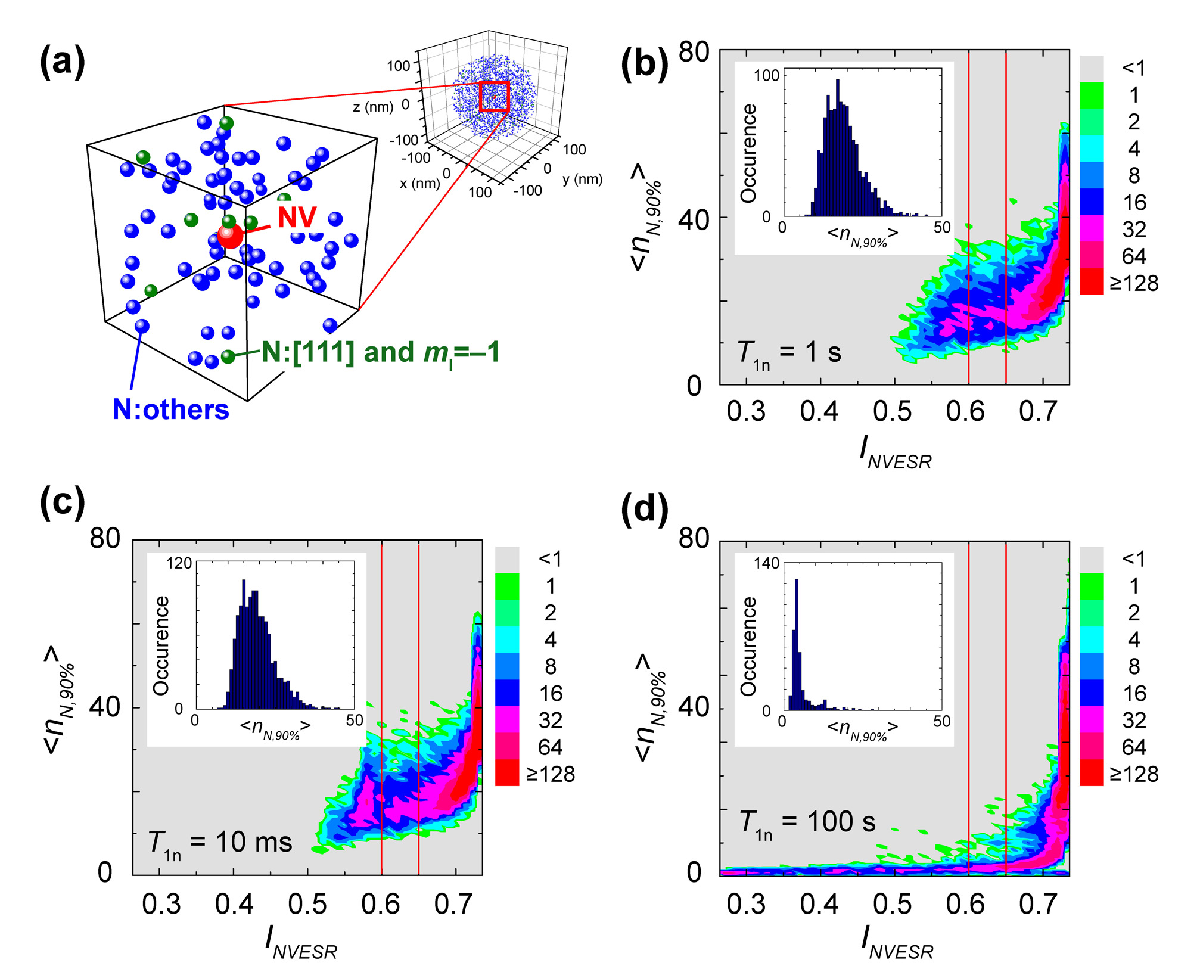}
\caption{
\label{fig:Sim}
Simulation of the NV-based ESR signals.
(a) Overview of the simulation model.
Diameter of the simulated lattice is $600a$ where $a=0.375$ nm is the lattice constant of diamond.
Red sphere denotes the NV center placed at the origin of the lattice.
Blue spheres represent randomly distributed N spins with 20 ppm concentration.
Green spheres are a subset of N spins which are in the $\langle$111$\rangle$ orientation with $m_I=-1$.
N spins at or next to the origin, next to other N spins, and overlap to other N spins, were suppressed in the simulation.
(b-d) Intensity plots of $\langle n_{N,90\%} \rangle$ versus $I_{NVESR}$ for $T_{1n}=1$ s, 10 ms, and 100 s, respectively.
$10^{4}$ spatial configurations were simulated for each $T_{1n}$.
The color scheme for intensity is shown in the legend.
Vertical solid lines indicate the intensity of the observed ESR signal, namely $I_{NVESR}$ = 0.60-0.65.
The insets show the occurrence of spatial configurations of the simulated lattice that resulted in  $0.60 < I_{NVESR} < 0.65$ as a function of $\langle n_{N,90\%} \rangle$.
The number of such configurations were $\sim$1200, $\sim$1300, and $\sim400$ for $T_{1n}=1$ s, 10 ms, and 100 s, respectively, which translate to 4$-$13 $\%$ out of total $10^4$ configurations.
}
\end{figure}
Finally, we analyze the intensity of the observed NV-based ESR signal to estimate the number of the detected N spins.
Our analysis here focuses on the signal at $f_{MW2}=0.90$ GHz corresponds to the $|m_S=-1/2, m_I=-1\rangle \leftrightarrow |m_S=1/2, m_I=-1\rangle$ transition of N spins oriented along the $\langle$111$\rangle$ direction (Fig.~\ref{fig:DEER}(b)).
In the analysis, we simulate DEER intensities by calculating the effective magnetic dipole field at the NV center from the surrounding N spins ($B_{dip,eff}$).
First step of the simulation was to generate a model configuration of the NV center and N spins in a diamond lattice, and this was done by placing the NV center at the origin of the diamond lattice and assigning the positions of N spins randomly in the lattice sites (see Fig.~\ref{fig:Sim}(a)).
Based on the $T_2$ measurement ($T_2$ = 1.2 $\mu$s $\sim$ 20 ppm), $\sim$18,000 N spins were placed in the simulated diamond lattice with a diameter of $\sim$200 nm.
Next, we randomly chose the orientation of N spins where only a quarter of N spins were assigned along the $\langle$111$\rangle$ direction and the rest of N spins were assigned along the other three directions, {\it i.e.}, $[\bar{1}11]$, $[1\bar{1}1]$, and $[11\bar{1}]$.
We then assigned the nuclear spin value ($m_I$) of either $1$, $0$, or $-1$ to all N spins with equal probability.
For the simulation, we only considered the contribution coming from the N spins oriented along the $\langle$111$\rangle$ direction with $m_I=-1$ (The signal at $f_{MW2}=0.90$ GHz in Fig.~\ref{fig:DEER}(b)).
Thus, $B_{dip,eff}$ was computed from only $1/12$-th of all N spins in the simulated lattice on average ($\sim$1500 N spins).
Finally we assigned the electron spin value ($m_S$) of either $1/2$ or $-1/2$ with equal probability to the N spins.

$B_{dip,eff}$ is given by the sum of individual dipolar field from each N spin as,
\begin{eqnarray}
B_{dip,eff} = \sum_{i=1}^{n_N} B_{dip,i}
\end{eqnarray}
where $B_{dip,i}=\frac{\mu_0}{4\pi} \frac{(3\cos^2 \theta_i - 1)}{r_i^3} g^N \mu_B m_S^i$ is the dipolar field strength of $i$-th N spin at the NV center, $n_N$ is the number of the N spins oriented along the $\langle$111$\rangle$ direction with $m_I=-1$, $\mu_0$ is the vacuum permeability, $r_i$ and $\theta_i$ are the magnitude and polar angle of the vector that connects the NV center and $i$-th N spin, respectively, and $m_S^i$ is the electron spin value of $i$-th N spin.
The mutual flip-flops within N electron spins was also not considered because of a low rate of the spin flip-flop process during the measurement, {\it i.e.} $\tau_C \gg \tau$ (see Fig.~\ref{fig:ODMR}(e)).
We then calculated the number of N spins ($n_{N,90\%}$) out of $n_N$ to obtain more than 90$\%$ of the $B_{dip,eff}$ by successively adding individual dipolar field term from each N spin in descending order of magnitude ({\it i.e.}, when $|B_{dip,eff}-\sum_{j=1}^{n_{N,90\%}} B_{dip,j}|/|B_{dip,eff}| \leq 0.1$).
In addition, we took into account for the electron and nuclear spin relaxations of N spins during the DEER measurement time ($\sim 100$ s with $\sim 10^7$ averaging) by statistically re-assigning the electron and nuclear spin values ($m_S$ and $m_I$) according to the electron and nuclear longitudinal relaxation times ($T_{1e}$ and $T_{1n}$), respectively.
Although $T_{1e}$ has been investigated previously ($T_{1e}$$\sim$10 ms),\cite{Reynhardt98, Takahashi08} $T_{1n}$ has not been reported before to the best of our knowledge.
In the simulation, we used $T_{1e}=10$ ms and considered $T_{1n}$ to be 10 ms, 1 s and 100 s, therefore the assignment of $m_I$ happens $10^2$ times for $T_{1n}=1$ s ($10^4$ and 1 times for $T_{1n}=10$ ms and 100 s, respectively) and a total of $10^4$ iterations (100 s/$T_{1e}$) occurs during the simulation.
Then, what we finally computed were the average of $10^4$ values of $n_{N,90\%}$ and $I_{DEER}$ ({\it i.e.}, $\langle n_{N,90\%} \rangle=\sum_{k=1}^{10^4} n_{N,90\%}^k$ and $\langle I_{DEER} \rangle=\sum_{k=1}^{10^4} I_{DEER}^k$),
therefore, by rewriting Eq.~(\ref{eq:NV_ESR}), the simulated NV-based ESR signal is given by,
\begin{eqnarray}
\label{eq:NV_ESR_P0}
I_{NVESR} = 1/2(1+ \langle I_{DEER} \rangle SE(2\tau)).
\end{eqnarray}
In addition, in order to see the spatial configuration dependence on $\langle n_{N,90\%} \rangle$ and $I_{NVESR}$,
we repeated the procedure described above for $10^4$ spatial configurations and calculated $10^4$ values of $\langle n_{N,90\%} \rangle$ and $I_{NVESR}$.

Figure~\ref{fig:Sim}(b)-(d) shows the simulated result of $\langle n_{N,90\%} \rangle$ as a function of $I_{NVESR}$ (Eq.~(\ref{eq:NV_ESR_P0})) for $T_{1n}=1$ s, 10 ms, and 100 s, respectively, from $10^4$ spatial configurations.
As shown in Fig.~\ref{fig:Sim}(b), the number of the N spins that contribute to the NV-based ESR signal ($\langle n_{N,90\%} \rangle$) depends on the detected $I_{NVESR}$.
We first noted that, for all $10^4$ configurations we simulated, only a small portion ($\leq$60) out of $\sim$1500 N spins contributes to more than 90 $\%$ of $B_{dip,eff}$ on average.
In the case of homogeneously distributed N spins, the 60 spins are located in a sphere with $\sim$70 nm diameter.
Moreover, when the NV-based ESR intensity is large ($I_{NVESR}\sim0.5$), $\langle n_{N,90\%} \rangle$ is smaller because $B_{dip,eff}$ in such spatial configurations is only dominated by a smaller ensemble of the N spin located in the vicinity of the NV center.
In addition, there was little observation of $\langle I_{DEER} \rangle<0$ ({\it i.e.}, $I_{NVESR}<0.5)$ because a small probability exists for the NV center to be coupled with a single or a few N spins with the 20 ppm N concentration and the resultant $\langle I_{DEER} \rangle$ is a weighted sum of many oscillatory functions with different frequencies.
On the other hand, when the NV-based ESR intensity is small ($\langle I_{DEER} \rangle\sim1$) ({\it i.e.} $I_{NVESR}\sim$ 0.73), $\langle n_{N,90\%} \rangle$ is larger because the N spins in such spatial configurations spread uniformly and the N spins located farther away from the NV also contribute to $B_{dip,eff}$.
The experimentally observed NV-based ESR intensity was $0.625\pm0.025$ (see Fig.~\ref{fig:DEER}(b)).
The inset of Fig.~\ref{fig:Sim}(b) shows a histogram of the occurrence of $\langle n_{N,90\%} \rangle$ from the simulation that yielded $I_{NVESR}=0.6$-0.65.
The occurrence was in the range of $7-42$ for $T_{1n}=1$ s.
Moreover, as shown in Fig.~\ref{fig:Sim}(c), $\langle n_{N,90\%} \rangle$ with $T_{1n}=10$ ms was similar to the result with $T_{1n}=1$ s (The occurrence was in the range of $7-45$).
On the other hand, in the case of $T_{1n}=100$ s, the distribution was slightly different and the occurrence was in $2-28$ as shown in Fig.~\ref{fig:Sim}(d).
Thus, from the simulation, the number of N spins in the present NV-based ESR measurement is estimated to be $\leq$ 50 spins.

%
% Summary
%
\section{Summary}
In summary, we presented ESR spectroscopy using a single NV center in diamond.
First, we demonstrated the identification of microscopic spin baths surrounding a single NV center and the investigation of static and dynamic properties of the bath spins using Rabi, FID, SE measurements as well as NV-based ESR spectroscopy.
We also performed the investigation with several other single NV centers in the diamond sample and
found that the properties of the bath spins are unique to the NV centers.
Finally, by analyzing the intensity of the NV-based ESR signal using the computer simulation,
we estimated the detected spins in the DEER measurement to be $\leq$ 50 spins.

%
% Acknowledgement
%
\section{Acknowledgements}
This work was supported by the National Science Foundation (DMR-1508661), the USC Anton B. Burg Foundation and the Searle scholars program (S.T.).

%
% References
%
%\bibliography{TakaLab}

%merlin.mbs aipnum4-1.bst 2010-07-25 4.21a (PWD, AO, DPC) hacked
%Control: key (0)
%Control: author (8) initials jnrlst
%Control: editor formatted (1) identically to author
%Control: production of article title (0) allowed
%Control: page (1) range
%Control: year (1) truncated
%Control: production of eprint (0) enabled
%

\end{document}